# Speech based Password Protected Cyber Applications

Ms. Urmila Shrawankar
RTM Nagpur University
Nagpur, India
urmilas@rediffmail.com
Cell : +919422803996

Dr. V. M. Thakre
SGB Amravati University
Amravati, India

*ABSTRACT*

Whenever we think of cyber applications, we visualize the model that gives the idea that we are sitting in front of computer at home or workplace connected to internet and performing all the work that generally we have to go and do on a specific place for example e-shopping, e-banking, e-education etc.

In case of e-shopping, we view all the products on the computer screen with all details by a single mouse click, select the product and do all further money transaction through net.

When we think of security, is it 100% secure? No, not at all because though it is password protected, the password is a text base secrete code that can be open.

Therefore this paper is concentrated on preparation of speech based password protected applications.

## 1   Introduction

The theme of this project is, while running any cyber application, whenever user / password is asked, user will provide the user name and password in the form of wav file (Voice/Speech format) through microphone.

First user will provide the sample passwords to train the machine. The password is in speech format produced by a speaker, i.e. the application is under category of speaker dependent, text independent, real-time speech and speaker recognition system.

The Hidden Morkov Model is used to train and test. Feature extration part is handled by signal processing front-end i.e. MFCC.

This model is strongly based on the biological voice production system. The anatomy of vice production system differs person to person and hence the unique voice is produce by every person.

A software is prepared for training the machine to recognize the uniqueness in person's voice.

I tried for implementing the features of text independent speech recognition system that will extract the features from persons voice for any text that will help to change the password frequently.

## 2   Speech Recognition System

When an e-application asks password, the credit card number or any number the user will speak the secrete text is his/her own voice, machine will extract the features from the speech input using digital signal processing front-end and pass the input for testing, if speech password/number verified the money transaction permitted else protect from the cyber crime.

### 2.1   How does Speech Recognition system work?

Speech Recognition is a technology, which allows control of machines by voice in the form of isolated or connected word sequences. It involves the recognition and understanding of spoken language by machine. One part of the process is recognizing the words that have been spoken without necessarily interpreting their meanings.

The other part is speech understanding in which the meaning is of the speech is ascertained.

The information in speech signal is actually represented by short-term amplitude spectrum of the speech waveform. This allows us to extract features based on the short-term amplitude spectrum from speech and to confidently use these features as the basic of pattern matching.

Speech Recognition is fundamentally a pattern classification task. The objective is to take an input pattern, the speech signal and classify it as a sequence of stored patterns that have precisely been defined. These stored patterns may be made of units, which we call *phonemes*.

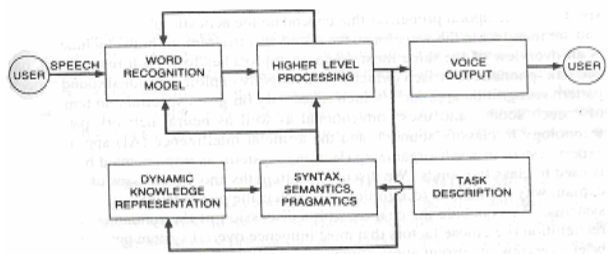

**General Block Diagram Of A Task-Oriented Speech Recognition System**

The task is to determine which of the variations in the speech are relevant to speech recognition and which variations are not relevant.

Based on the difference in the way, words are pronounced, there are three standard modes of speaking to a machine, namely;

- Isolated Word Recognition
- Connected Word Recognition
- Speaker Dependent
- Speaker Independent
- Text Dependent
- Text Independent

This project is based on *Isolated Word Recognition*, *Speaker Dependent* and *Text Independent* model.

## 2.2 Approaches to Automatic Speech Recognition by Machine

The Speech recognition by the machine, whereby the machine attempts to decode the speech signal in a sequential manner based on the observed acoustic features of the signal and the known relations between acoustic features and phonetic symbols.

**The Artificial Intelligence Approach**

The artificial intelligence approach to speech recognition is a hybrid to the acoustic phonetic approach and the pattern recognition approach in that it exploits ideas and concepts of both methods. The artificial intelligence approach attempts to mechanize the recognition procedure according to the way a person applies its intelligence in visualizing analyzing, and finally making a decision on the measures acoustic features. The use of neural networks for learning the relationships between phonetic events and all known inputs, as well as for discrimination between similar sound classes.

## 3. Feature extractions And feature matching

Feature extraction is the process that extracts a small amount of data from the voice that can later be used to represent each speaker. Feature matching involves the actual procedure to identify the unknown speaker by comparing extracted features from his/her voice input with the ones from a set of known speakers.
All speaker recognition systems have to serve two distinguishes phases. The first one is referred to the enrollment sessions or training phase while the other is referred to as the operation sessions or testing phase.
In the training phase, each registered speaker has to provide samples of their speech so that the system can train a reference model for that speaker.

In the case of speaker verification systems, in addition, a speaker specific threshold is also computed from the training samples. During the testing phase, the input speech is matched with stored reference model(s) and recognition decision is made.

### 3.1 Speech Feature Extraction

The purpose of this module is to convert the speech waveform to some type of parametric representation (at a considerably lower information rate) for further analysis and processing. This is often referred as the *signal-processing front end*.

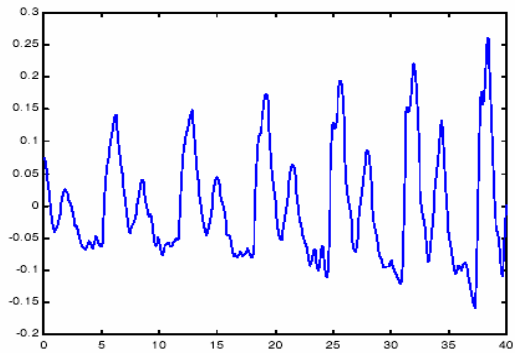

**An example of speech signal**

A wide range of possibilities exist for parametrically representing the speech signal for parametrically representing the speech signal and the speaker recognition task, such as Linear Prediction Coding (LPC), Mel-Frequency Cepstrum Coefficients (MFCC), and others MFCC's are based on the known variation of the human ear's critical bandwidths with frequency, filters spaced linearly at low frequencies and logarithmically at high frequencies have been used to capture the phonetically important characteristics of speech. This is expressed in the *mel-frequency* scale, which is a linear frequency spacing below 1000 Hz and a logarithmic spacing above 1000 Hz.

### 3.2 Mel-Frequency Cepstrum Coefficients (MFCC) Model

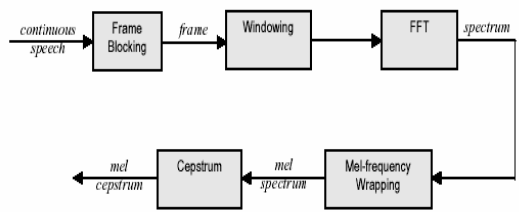

**MFCC processor**

A block diagram of the structure of an MFCC processor is given in Figure mfcc. The speech input is typically recorded at a sampling rate above 10000 Hz. This sampling frequency was chosen to minimize the effects of aliasing in the analog-to-digital conversion. These sampled signals can capture all frequencies up to 5 kHz, which cover most energy of sounds that are generated by humans.

### 3.3 Training with Hidden Markov Model

In the context of statistical methods for speech recognition, hidden Markov models (HMM) have become a well known and widely used statistical approach to characterizing the spectral properties of frames of speech. As a stochastic modeling tool, HMMs have an advantage of providing a natural and highly reliable way of recognizing speech for a wide variety of applications. In HMM the observed data are viewed as the result of having passed the true (hidden) process through a function that produces the second process (observed). The hidden process consists of a collection of states (which are presumed abstractly to correspond to states of the speech production process) connected by transitions. Each transition is described by two sets of probabilities:

• A **transition probability**, which provides the probability of making a transition from one state to another.
• An **output probability** density function, which defines the conditional probability of observing
The continuous density function most frequently used for this purpose is the multivariate Gaussian mixture density function
The goal of the decoding (or recognition) process in HMMs is to determine a sequence of (hidden) states (or transitions) that the observed signal has gone through. The second goal is to define the likelihood of observing that particular event given a state determined in the first process.

### 3.4 Recognition

An HMM can be used to model a specific unit of speech. The specific unit of speech can be a word, a subword, or a complete sentence or paragraph. In large-vocabulary systems, HMMs are usually used to model subword units such as phonemes, while in small-vocabulary systems HMMs tend to be used to model the words themselves.
The training procedure involves optimizing HMM parameters given an ensemble of training data.

An iterative procedure, the Baum-Welch or forward-backward algorithm, is employed to estimate transition probabilities, output distributions, and codebook means and variances under a unified probabilistic framework. Viterbi algorithm is used as a fast-match algorithm

## 4. Adverse Conditions In SR System

While developing this project, it is observed that some adverse conditions degrade the performance of the Speech Recognition system.

### 4.1 Noise
If we use the noise free room to train and test both the time we are getting about 80% accuracy. But if the room is noisy either in training phase or in testing phase accuracy is reduces to around 60%

### 4.2 Distortion
To implement this project we do not require any special hardware than the computer machine, (where we run the cyber application), a Microphone to provided secrete code / password and a headphone to hear the echo of password provided. If these attachments and not installed and configure properly we get distorted input signals, which reduces the accuracy.

### 4.3 (Human) Articulation Effects
Many factors affect the manner of speaking of each individual talker, like the distance of microphone from the speaker and its position also speech added with psychological effect while providing and the password these factors effects the accuracy

## 5. Conclusion

After implementing this software project for any cybernetic application user may get more security in cyber space. Person's unique voice features will help for protecting password.

These types of application will be more beneficial to handicap or old-aged persons those are unable to do movement or not able to operate the keyboard.